\documentstyle[psfig,aaspp4]{article}

\def\etal{et al.\,}
\def\Bruntfreq{Brunt-V{\"a}is{\"a}l{\"a}\,\,}
\def\refnew#1{(\ref{#1})}
\def\be{\begin{equation}}
\def\ee{\end{equation}}

\def\vcv{v_{\rm cv}}
\def\tcv{t_{\rm cv}}

\def\K{\, \rm K}
\def\s{\, \rm s}
\def\km{\, \rm km}
\def\cm{\, \rm cm}
\def\dyne{\, \rm dyne}

\def\be{\begin{equation}}
\def\ee{\end{equation}}

\begin{document} 

\title{\mbox{GRAVITY-MODES IN ZZ CETI STARS}\\ 
\mbox {III. EIGENVALUES AND EIGENFUCTIONS}}
\author{Yanqin Wu\altaffilmark{1,2} and Peter Goldreich\altaffilmark{1}}

\altaffiltext{1}{Theoretical Astrophysics, California Institute of Technology 
	130-33, Pasadena, CA 91125, USA; pmg@gps.caltech.edu}
\altaffiltext{2}{Astronomy Unit, School of Mathematical Sciences, 
	 Queen Mary and Westfield College, Mile End Road, London E1 4NS, UK;
                 Y.Wu@qmw.ac.uk}

\begin{abstract} 
We report on numerical calculations of nonadiabatic eigenvalues and
eigenfunctions for g-modes in ZZ Ceti variables. The spectrum of
overstable $\ell=1$ modes delineates the instability strip. Its blue
edge occurs where $\omega \tau_c \approx 1$ for the $n=1$ mode. Here
$\omega$ is radian frequency and $\tau_c$ is about four times the
thermal timescale at the bottom of the surface convection zone. As a
ZZ Ceti cools, its convection zone deepens, longer period modes become
overstable, but the critical value of $\omega\tau_c$ separating
overstable and damped modes rises. The latter is a consequence of
enhanced radiative damping for modes which propagate immediately below
the convection zone. The critical value of $\omega\tau_c$ is of
observational significance because modes with the smallest value of
$\omega\tau_c$ are most observable photometrically.  Maximum periods
for overstable modes predicted for our cooler model envelopes are
about a factor two longer than the observational upper limit of
$1,\!200\s$. We assess a number of plausible resolutions for this
discrepancy among which convective overshoot and nonlinear saturation
look promising.  The nonadiabatic eigenfunctions enable us to predict
relative amplitudes and phases of photospheric variations of flux and
velocity, quantities made accessible by recent observations. We also
present asymptotic formula for damping rates of high order modes, a
result of consequence for future investigations of nonlinear
saturation of the amplidues of overstable modes.
\end{abstract}

\section{Introduction}
\label{sec:nonad-intro}

While passing through a narrow temperature range around $T_{\rm eff}
\sim 12,\!000 \K$, DA (hydrogen) white dwarfs 
exhibit pulsations with periods between 2 and 20 minutes. These
oscillations are identified with g-modes having radial orders $1\leq
n\leq 30$, and spherical harmonic degrees $1\leq
\ell\leq 2$.  Mode overstability occurs when driving in the thin and
fast reacting surface convection zone exceeds damping in the
underlying radiative interior (Brickhill
\cite{nonad-brick91}, Gautschy et al. \cite{nonad-gautschy96}, Goldreich \& Wu 
\cite{nonad-paper1}, hereafter Paper I). The flux perturbation entering the 
convection zone from the underlying radiative interior is in phase
with the pressure perturbation.  Because the convection is fast and
efficient, the specific entropy perturbation is nearly
depth-independent throughout the bulk of the convection
zone. Consequently, the magnitude of the flux perturbation decreases
outward. The phase relation between the pressure and flux
perturbations then results in convective driving.

Gravity-modes have simple structures in their upper evanescent
regions. This permits an analytic derivation of a stability criterion
based on the quasiadiabatic approximation for modes whose evanescent
regions extend well below the base of the convection
`zone. Overstability is predicted provided $\omega \tau_c > 1$,
where $\tau_c$ is a few times the thermal time scale at the bottom of
the convection zone. Applied blindly, this criterion predicts
overstability in cool ZZ Cetis for modes whose periods far exceed
those observed. Our numerical calculations enable us to drop both the
quasiadiabatic approximation and the restriction that the evanescent
region extend well below the convection zone. They also allow us to
incorporate the effect of turbulent viscosity as discussed in
Brickhill (\cite{nonad-brick90}) and Goldreich \& Wu
(\cite{nonad-paper2}, hereafter Paper II).

This paper is the third in a series dedicated to the overstability of
gravity-modes in ZZ Cetis. It is largely numerical, and presents
results of nonadiabatic calculations of eigenvalues and
eigenfunctions. Nonadiabaticity is important for modes which have
$\omega \tau_c\lesssim 1$, since they have $\tau_{\rm th}< 1$ in their
driving and damping regions. This category includes both overstable
modes with $\omega \tau_c\approx 1$, which are easily detectable by
photometric measurements, and damped modes with $\omega \tau_c\ll 1$,
whose excitation by parametric instability acts to limit the
amplitudes of overstable modes. It is imperative that our calculations
accurately represent them.

The organization of this paper is as follows.  In \S
\ref{sec:nonad-eqsetup}, we collect the equations which govern linear,
nonadiabatic oscillations, establish appropriate boundary conditions,
and describe the method we use to construct envelope models for ZZ
Cetis.  In \S \ref{sec:nonad-important} we assess the magnitude of
nonadiabatic effects in the evanescent and propagating regions of
g-modes. Numerical techniques employed to solve the eigenvalue problem
are elaborated in \S \ref{sec:nonad-numer}.  The impact of
nonadiabaticity and turbulent viscosity on g-mode eigenvalues and
eigenfunctions is described in \S \ref{sec:nonad-result}. We then
discuss in \S \ref{sec:nonad-maximum} several possibilities to
reconcile the nonadiabatically calculated spectra of overstable modes
with those obesrved.  A short summary follows in \S
\ref{sec:nonad-sum}. The appendix is devoted to analysis of a toy
model which elucidates the asymptotic damping rates of high order
modes.

Symbols used in this paper are defined in Table 1. Their usage is consistent
with that in previous papers of this series.

\begin{deluxetable}{cll}[t]
\tablewidth{0pc}
\tablecaption{DEFINITIONS}
\tablehead{
\colhead{Symbol}      & \colhead{Meaning}} 
\startdata
$R$ & stellar radius \nl
$g$ & surface gravity \nl
$z$ & depth below photosphere \nl
$z_b$ & depth at bottom of convection zone \nl
$z_\omega$ & depth at top of mode's cavity,\, $z_\omega\sim \omega^2/(gk_h^2)$ 
\nl
$\omega$ & radian mode frequency, $\omega = \omega_r + i \omega_i$\nl
$n$ & radial order of mode  \nl
$\ell$ & angular degree of mode \nl
$k_h$ & horizontal wave vector,\,  $k_h^2=\ell(\ell+1)/R^2$ \nl
$k_z$ & vertical wave vector \nl
$\rho$ & mass density, $\rho_b$ mass density at $z= z_b$ \nl
$p$ & pressure \nl
$T$ & temperature \nl
$s$ & specific entropy in units of $k_B/m_p$ \nl
$F$ & energy flux \nl
$c_s$ & adiabatic sound speed,\,  $c_s^2=(\partial p/\partial \rho)_s = \Gamma_1 
p/\rho$ \nl
$\rho_s$ & $\rho_s\equiv (\partial\ln\rho/\partial s)_p$ \nl
$\delta$ & denotes Lagrangian perturbation \nl
$\xi_h$ & horizontal component of displacement vector  \nl
$\xi_z$ & vertical component of displacement vector \nl
$v_h$ & horizontal component of velocity vector \nl
$f$ & coefficient measuring convective inefficiency \nl
$\vcv$ & convective velocity,\,  $\vcv\sim (F/\rho)^{1/3}$ \nl
$\tcv$ & response time for convection,\,  $\tcv\sim z/\vcv$ \nl
$A, \,\, B, \,\, C$ & dimensionless constants approximately $2, \,\, 8$ \& $8$
for ZZ Cetis \nl
$\tau_{\rm th}$ & thermal constant at depth z,\, $\tcv/\tau_{\rm th}\sim
(\vcv/c_s)^2$ in the convection zone \nl
$\tau_{\omega}$ & thermal time at $z = z_\omega$ \nl
$\tau_b$ & unconventional thermal time constant at $z_b$,\, $\tau_b\approx 
\tau_{\rm th}/5$ at $z_b$.  \nl
$\tau_c$ & time constant of low pass filter for convection zone, \, 
$\tau_c=(B+C)\tau_b$ \nl 
$N^2$ & \Bruntfreq frequency\nl
$\kappa_T$, $\kappa_\rho$ & opacity derivatives, $\kappa_T \equiv (\partial \ln 
\kappa/\partial \ln T)_\rho$, 
$\kappa_\rho \equiv (\partial \ln \kappa/\partial \ln \rho)_T$ \nl
$c_p$ & dimensionless heat capacity, $c_p \equiv (\partial s/\partial \ln T)_p$ 
\nl
$\nabla$ & temperature gradient, $\nabla \equiv d\ln T/d\ln p)$ \nl
$\nabla_{\rm ad}$ & adiabatic temperature gradient, $\nabla_{\rm ad}
\equiv (\partial \ln T/\partial \ln p)_{s}$ \nl
\enddata
\end{deluxetable}

\section{Equations and Boundary Conditions}
\label{sec:nonad-eqsetup}

\subsection{Equations in the Radiative Region}
\label{subsec:nonad-eqnrad}

The linearized equations describing nonadiabatic pulsations read
\begin{eqnarray}
{{\delta \rho}\over\rho} & = &
 - i k_h \xi_h -  {d\xi_z\over{dz}}, \label{eq:nonad-cont} \\ 
{{\omega^2}\over g} \xi_h & = & i k_h \left( {p\over {g \rho}}
{{\delta p}\over p} -  \xi_z \right), \label{eq:nonad-xih} \\ 
{{\omega^2}\over g} \xi_z & = & {p\over{\rho
g}}{d\over{dz}}\left({{\delta p}\over p}\right) - 
 {d\xi_z\over{dz}}  
+ {{\delta p}\over p} - {{\delta \rho}\over \rho} ,
\label{eq:nonad-xiz} \\ 
\delta s & = & {iF m_p\over \omega \rho k_B T}
{d\over{dz}} \left({{\delta F}\over F}\right), 
\label{eq:nonad-s} \\ 
{{\delta F}\over F} & = & (4-\kappa_T){{\delta T}\over T} -
(1+\kappa_\rho) {{\delta \rho}\over \rho} - {{d\xi_z}\over{dz}} +
{p\over{\rho g\nabla}} {d\over{dz}}
\left({{\delta T}\over T}\right),
\label{eq:nonad-delF}
\end{eqnarray}
for plane-parallel geometry with the Cowling approximation, and an
assumed time-dependence of $\exp^{-i \omega t}$.  Equations
\refnew{eq:nonad-cont}-\refnew{eq:nonad-s} express the conservation of
mass, horizontal and vertical momentum, as well as energy.  Equation
\refnew{eq:nonad-delF} describes the equation of radiative
diffusion. Closure of this system of equations requires constitutive
relations for the equation of state and the opacity. Setting $\delta
s=0$ reduces the above equations to the adiabatic ones studied in
Paper I.

We choose $\delta p$ and $\delta s$ as our independent thermodynamic variables
and set
\begin{eqnarray}
{{\delta \rho}\over \rho} & = & {1\over \Gamma_1} {{\delta p}\over p} +
\rho_s \delta s, \label{eq:nonad-thermorho} \\
{{\delta T}\over T} & = & \nabla_{ad} {{\delta p}\over p} + {\delta s\over c_p}.
\label{eq:nonad-thermoT}
\end{eqnarray}
Then the linear perturbation equations may be written as four first-order 
differential equations for the four dependent variables $\delta p/p$, 
$d(\delta p/p)/d\ln p$, $\delta F/F$, and $\delta s$; 
\begin{eqnarray}
{d\over d\ln p}\left({{\delta p}\over p}\right) & = & 
 X, \label{eq:nonad-defineX} \\
{dX\over d\ln p} & = & 
-\left({p\over{g\rho}}\right)^2
\left[ k_h^2 \left({N^2\over \omega^2} -1\right) 
+\left({\omega \over {c_s}}\right)^2 \right]\left({{\delta p}\over p}\right) 
-X +\left({p\over {g\rho}}\right) {{(gk_h)^2-\omega^4}\over{g \omega^2}} \rho_s
\delta s, 
\label{eq:nonad-dppds} \\
{d\over d\ln p} \left({{\delta F}\over F}\right) & = & 
{-i\omega k_B T p\over g m_p F} \delta s,
\label{eq:nonad-dsdFF} \\
 {1\over c_p \nabla}{d\delta
 s\over d\ln p} & = &  - \left[ (4-\kappa_T) \nabla_{ad} - 
  {{\kappa_\rho}\over \Gamma_1}  
  -\left( 1- {{\omega^2 p }\over{g^2\rho }}\right) 
  {{(gk_h)^2}\over{(gk_h)^2-\omega^4}} + 
  {1\over \nabla} {{d \nabla_{ad}}\over{d\ln p}}\right]
  \left({{\delta p}\over p}\right) 
\nonumber \\
& &\hskip-0.1in -\left[ {{\nabla_{ad}}\over{\nabla}} -
   {{(gk_h)^2}\over{(gk_h)^2-\omega^4}} \right] X
   +\left({{\delta F}\over F}\right) 
 - \left[ {{(4-\kappa_T)}\over{c_p}} -\kappa_\rho \rho_s + {1\over\nabla}
{{d}\over {d\ln p}}\left({1\over {c_p}}\right) \right] \delta s.
\label{eq:nonad-dffall} 
\end{eqnarray} 

The perturbation equations are to be solved as an eigenvalue problem. Since 
equation \refnew{eq:nonad-dsdFF} includes a factor $i$, all four dependent 
variables and the eigenfrequency are {\it complex}. We write each complex 
variable in the form
\begin{equation}
Q = ( Q_r + i Q_i ) \exp^{- i \omega t}.
\label{eq:nonad-complex}
\end{equation}
The physical perturbation is given by ${\rm Re}(Q)=e^{\omega_i t} (Q_r 
\cos{\omega_r t} + Q_i \sin{\omega_r t})$.

\subsection{Equations in the Convection Zone}
\label{subsec:nonad-eqnconv}

Inside the convection zone, g-mode perturbations are constrained by
the rapid response of convection to the instantaneous pulsational
state (Brickhill \cite{nonad-brick90} \& \cite{nonad-brick91}, Papers
I \& II).  Rapid momentum diffusion enforces
\be
|X|=\left|{d\over d\ln p}\left({\delta p\over p}\right)\right|\ll {z\over 
z_\omega}\left|{\delta p\over p}\right|.
\label{eq:nonad-Xconst}
\ee
And fast entropy mixing ensures that
\be
\left|{d\delta s\over d\ln p}\right|\ll {z\over z_\omega}\left|\delta s\right|,
\label{eq:nonad-dscvz} 
\ee 
except in the thin superadiabatic layer. The entropy gradient in the
superadiabatic layer is non-negligible and increases with increasing
convective flux. But as this region is thin and has low mass, it is
sufficient to incorporate its effect into the boundary conditions.
The total, convective plus radiative, flux perturbation follows from
equation \refnew{eq:nonad-dsdFF}. whereas equation
\refnew{eq:nonad-dffall} determines the radiative flux perturbation in
terms of $\delta p$ and $\delta s$.

\subsection{Boundary Conditions} 
\label{subsec:nonad-bc}

Solving four linear, homogeneous, first-order, ordinary differential
equations to obtain eigenvalues and eigenfunctions, requires a total
of five boundary conditions. Four of these express physical
constraints imposed by the environment outside the domain in which the
differential equations are to be integrated. The fifth merely sets the
magnitude scale and phase for the eigenfunctions.

Approximations described in \S \ref{subsec:nonad-eqnconv} enable us to
lower the outer boundary from the photosphere to the top of the radiative 
interior at $z_b$. The three boundary conditions applied there read:
\begin{eqnarray}
\left({{\delta p}\over p}\right) & = & C,
\label{eq:nonad-dppzb} \\
X & = & {{-k_h^2}
\over{\omega^2 p_b}}\int_0^{p_b} dp\,{{p\rho_s}\over{\rho}} 
\left[{ds\over d\ln p}\left({{\delta p}\over p}\right) -
\delta s  \right],
\label{eq:nonad-ddppzb} \\
\delta s &  = & {(B+C)\over 1-i\omega\tau_c}\left({{\delta F}\over F}\right).
\label{eq:nonad-dszb}
\end{eqnarray}
The constant $C$ in equation \refnew{eq:nonad-dppzb} sets the scale
and phase of the eigenfunction. Equation \refnew{eq:nonad-ddppzb}
follows from the near vanishing of $X$ in the convection zone, as
expressed by equation
\refnew{eq:nonad-Xconst}, together with equations (20) and (21) of Paper II, 
which account for the jump in $X$ across the convective-radiative boundary. 
Equation \refnew{eq:nonad-dszb} relates the
entropy perturbation in the main part of the convection zone to the flux 
perturbation that enters from below (see Paper I). 

It is advantageous to raise the bottom boundary from the center of the
star to a depth, $z = z_{\rm deep}$, where the quasiadiabatic
approximation is valid and the plane-parallel approximation still
applies. This step alters the spectrum of eigenvalues
$\omega$. However, a simple procedure to be described later allows us
to recover values for $\omega_i$ appropriate to a complete stellar
model. We generally take $z = z_{\rm deep}$ to be the level at which
$p=10^{16}\dyne\cm^{-2}$. Since $\tau_{\rm th}\approx 10^{10}\s$ at
$z_{\rm deep}$, the quasiadiabatic approximation is valid there for
all g-modes of interest to our investigation. And in DA white dwarf
models provided to us by Bradley (\cite{nonad-bradley96}), the region
above $z_{\rm deep}$ extends over the outer $2$ percent of the stellar
radius and includes about $10^{-6}$ of the stellar mass. The two
boundary conditions imposed at $z = z_{\rm deep}$ are
\begin{eqnarray}
X + \left(1 - {{k_h^2 p}\over{\omega^2 \rho}}\right) {{\delta p}\over
p} & = & 0 ,
\label{eq:nonad-ddppzdeep} \\
\left({{\delta F}\over F}\right) + M_1 \left({{\delta p}\over
p}\right) + M_2 X & = & 0.
\label{eq:nonad-dffzdeep} 
\end{eqnarray}
The mechanical boundary condition given by equation \refnew{eq:nonad-ddppzdeep} 
states that $\xi_z=0$ at $z = z_{\rm deep}$. It follows from equations 
\refnew{eq:nonad-xih} and \refnew{eq:nonad-xiz}. Thus our model is bounded from 
below by a rigid wall. Equation \refnew{eq:nonad-dffzdeep} is our
thermal boundary condition. It is the quasiadiabatic limit of the
radiative diffusion equation \refnew{eq:nonad-dffall} which defines
the coefficients $M_1$ and $M_2$. Our lower boundary conditions are
somewhat arbitrary. For example, we could have adopted $d\xi_h/dz=0$
and $\delta s=0$ as mechanical and thermal boundary
conditions. However, because $\delta s$ decreases rapidly with depth,
taking $\delta s=0$ as the thermal boundary condition makes it more
difficult for our numerical schemes to converge on eigenvalues.

\begin{figure}[t]
\centerline{\psfig{figure=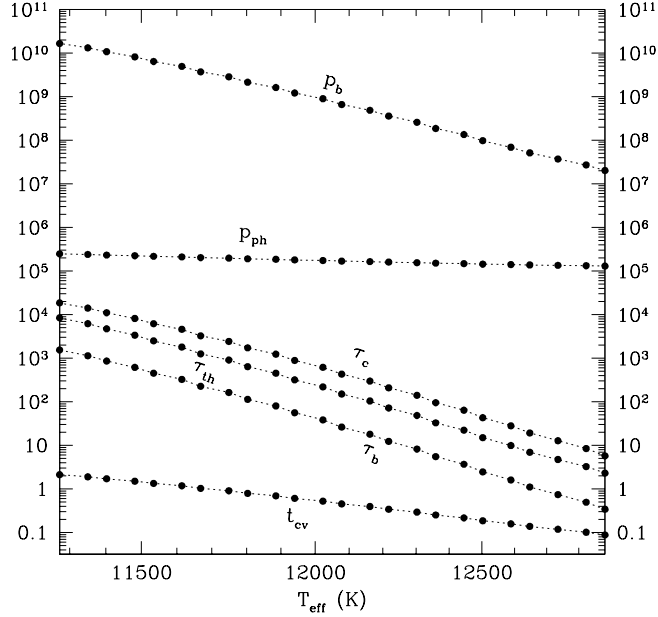,width=0.60\hsize}}
\caption[]{Convection zone characteristics for a sequence of hydrogen 
envelope models covering the temperature range where ZZ Cetis reside.
Models have surface gravity, $g=10^8\cm \s^{-2}$, and convection
parameter, $f=0.32$. Photospheric pressure ($p_{\rm ph}$) and pressure
at $z_b$($p_b$) are in units of$\dyne\cm^{-2}$.  Time constants
$t_{\rm cv}$, $\tau_b$, $\tau_{\rm th}$, and $\tau_c$ are in units of
$\s$. Both $t_{\rm cv}$ and $\tau_{\rm th}$ are evaluated at $z_b$.
Observationally detected g-modes have periods in the range $10^2-10^3
\s$.}
\label{fig:na-env-model}
\end{figure}

\subsection{Envelope Models}
\label{subsec:nonad-envelope}

Instead of using complete white dwarf models, we work with
plane-parallel, hydrogen envelopes computed on fine grids. These are
produced by integrating downward from the photosphere\footnote{The
photosphere is taken to be where $p = 2g/(3\kappa)$.} using the
Lawrence Livermore equation of state and opacity tables (Rogers \etal
\cite{nonad-rogers96}, Iglesias \& Rogers
\cite{nonad-iglesias96}). Our envelopes have $g=10^8\cm\s^{-2}$ and
cover the range $11,\!000\K \leq T_{\rm eff}\leq 13,\!000 \K$.

We model convection by invoking the mixing length ansatz. This involves
introducing a dimensionless parameter $f$ such that\footnote{See equation (23) 
of Paper I.} 
\be
{{ds}\over{d\ln p}}
= {f\over\left(|\rho_s| \rho p\right)^{1\over 3}} \left(
{{m_p}\over {k_B}} {{F_{\rm cv}}\over T}\right)^{2\over 3}.
\label{eq:nonad-cvzparameter}
\ee
The parameter $f$ is of order unity and is related to the conventional
mixing length ratio $\alpha$ by $f \sim \alpha^{-4/3}$. A smaller $f$
signifies more efficient convection and yields a thicker convection
zone. The radiative flux is related to the entropy gradient by
\be
{{ds}\over{d\ln p}} = 
{{3 \kappa p F_{\rm rad}}\over{16 \sigma T^4 g}} - \nabla_{ad} .
\label{eq:nonad-superparameter}
\ee
We determine $ds/d\ln p$, $F_{\rm cv}$, and $F_{\rm rad}$ from equations 
\refnew{eq:nonad-cvzparameter} and \refnew{eq:nonad-superparameter} together
with the relation $F=F_{\rm cv}+F_{\rm rad}$.

The grids of our model envelopes are chosen fine enough to resolve the
steep entropy profiles in the superadiabatic layer. Figure 
\ref{fig:na-env-model} displays some characteristics of
the surface convection zones in envelope models produced with
$f=0.32$. This value of $f$ enables us to match our model to that of
Bradley's at $T_{\rm eff} = 12,\!420 \K$.  Notice that the eddy
turn-over time, $\tcv$, is of order a few seconds in even the coolest
models of interest.

\begin{figure}[t]
\centerline{\psfig{figure=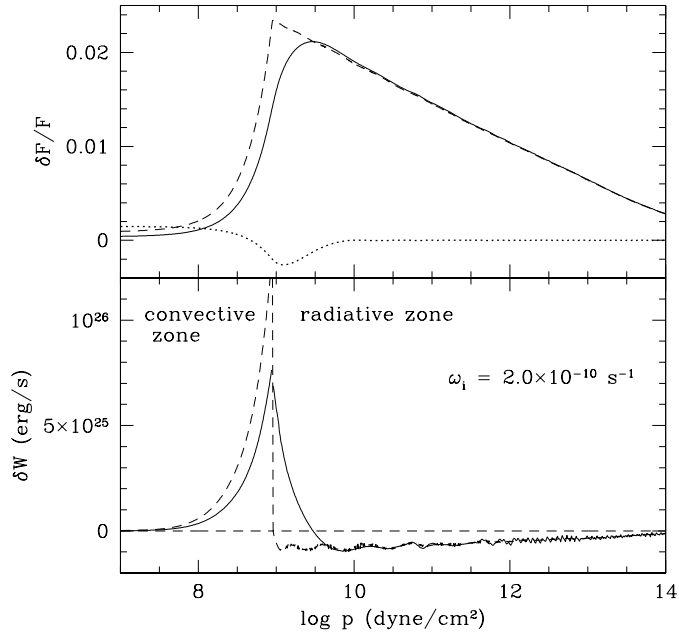,width=0.60\hsize}}
\caption[]{Comparisons of eigenfunctions calculated with and 
without the inclusion of radiative diffusion. The upper panel shows the
fractional Lagrangian flux perturbation, $\delta F/F$, and the lower the
differential work, defined by $W = \int \delta W d\log p$, both plotted against
$\log p$. Results from quasiadiabatic calculations are depicted by dashed lines,
whereas those from nonadiabatic calculations are given by solid lines for real
parts and dotted lines for imaginary parts. We set the nonadiabatic $\delta F/F$
to have the same phase and amplitude as the quasiadiabatic one in the deep
adiabatic interior.  The stellar model has $T_{\rm eff}=12,\!000\K$, which with
$g=10^8\cm\s^{-2}$ and $f=0.32$ yields $\tau_{\rm th}=200\s$ at $z_b$.  The mode
shown here has $\ell=1$ and $P=430 \s$. It is fairly adiabatic in the radiative
region since $\omega \tau_{\rm th} = 2.8$ at $z_b$. Nevertheless, radiative
diffusion noticebly reduces convective driving and gives rise to a small region
of radiative driving immediately below the convection zone. }
\label{fig:na-nad430}
\end{figure}

\begin{figure}[t]
\centerline{\psfig{figure=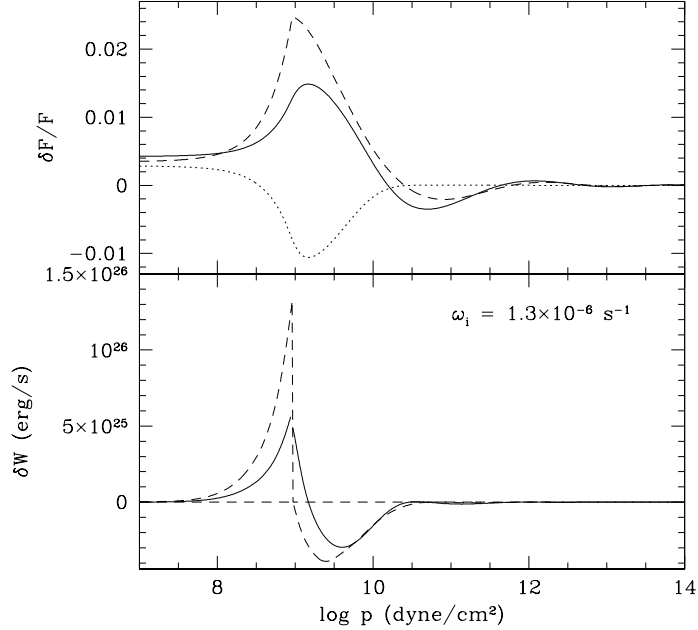,width=0.6\hsize}}
\caption[]{Similar to Figure \ref{fig:na-nad430}, but for a mode with $\ell=1$ 
and $P=1,\!500\s$. Since $\omega \tau_{\rm th} = 0.8$ at $z_b$, this mode is 
moderately nonadiabatic at the top of the radiative interior.
The mode remains overstable when radiative diffusion is taken into account.
However, the flux perturbation at the photosphere is modified in both 
amplitude and phase. }
\label{fig:na-nad1500-new}
\end{figure}

\begin{figure}[t]
\centerline{\psfig{figure=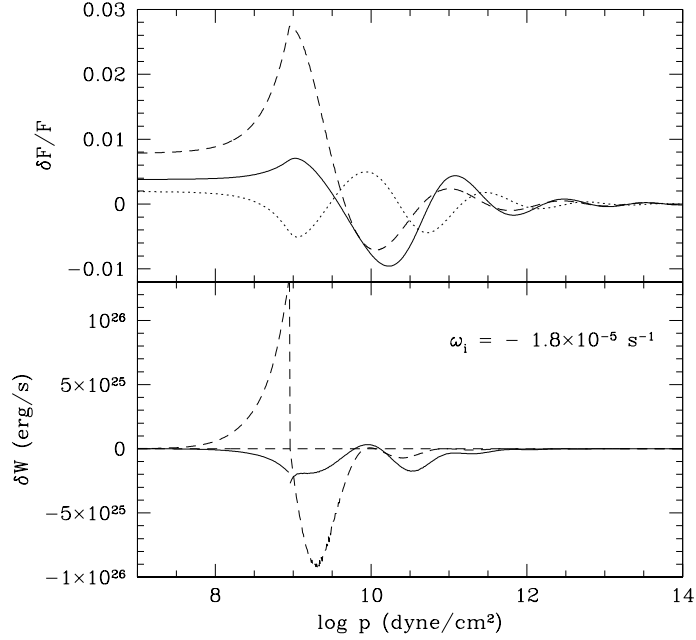,width=0.6\hsize}}
\caption[]{Similar to Figure \ref{fig:na-nad430}, but for a mode with $\ell=1$ 
and $P=2,\!400\s$ corresponding to $\omega_\tau{\rm th} = 0.5$ at
$z_b$. The nonadiabatic region extends down into the propagating
cavity. There is a large phase difference between $\delta F/F$ and
$\delta p/p$ at $z_b$ which leads to convective damping. As a
consequence, this mode is damped, although a quasiadiabatic
calculation predicts overstability.}
\label{fig:na-nad2400}
\end{figure}

\section{Where Nonadiabaticity Is Important}
\label{sec:nonad-important}

Nonadiabaticity is quantified by how much the presence of the entropy
perturbation, $\delta s$, affects $\delta p/p$ and $\delta F/F$.  Our
analysis for the radiative interior leads to separate criteria for the
evanescent region and the propagating cavity.

\subsection{Nonadiabatic Effects in the Evanescent Region}
\label{sec:nonad-evanescent}

The effect of $\delta s$ on $\delta p/p$ is contained in equation
\refnew{eq:nonad-dppds}. Noting that $\delta p/p$ varies on the scale 
$z_\omega > z$, we obtain
\begin{mathletters}
\begin{eqnarray}
 {d\over d\ln p}\left({{\delta p}\over p}\right)  &\sim &
{z\over{z_\omega}}\left({{\delta p}\over p}\right), \\
 {dX\over{d\ln p}}  & \sim &
{z\over{z_\omega}} \left({{\delta p}\over p}\right), \\
 \left({p\over{\rho g}}\right)^2
\left[ k_h^2 \left({N^2\over \omega^2} -1\right) + \left({\omega \over
{c^2}}\right)^2 \right] \left({{\delta p}\over p}\right) & \sim &
\left({k_h z N\over\omega}\right)^2\left({{\delta p}\over p}\right)
 \sim  {z\over{z_\omega}} \left({{\delta p}\over p}\right),
\label{eq:nonad-scalesp}
\end{eqnarray}
\end{mathletters}
where $z\sim p/(g\rho)$ and $N^2 \sim g/z$.
The nonadiabatic term is of order,
\be
{p\over{\rho g}} {{(gk_h)^2-\omega^4}\over{g \omega^2}} \rho_s
\delta s \sim  {z\over{z_\omega}} \delta s.
\label{eq:nonad-scalesds}
\ee
So the nonadiabatic correction to $\delta p/p$ is of order $\delta s$.

To relate $\delta s$ to $\delta p/p$, we turn to equations
\refnew{eq:nonad-dsdFF} and \refnew{eq:nonad-dffall}. With appropriate
scalings they yield\footnote{The coefficient connecting $\delta p/p$
to $\delta F/F$ in equation \refnew{eq:nonad-dffall} varies on scale
$z$, although weakly.}
\begin{mathletters}
\begin{eqnarray}
\delta s & \sim & {1\over{\omega \tau_{\rm th}}} 
{d\over{d\ln p}}\left({{\delta F}\over F}\right) 
\sim {1\over{\omega \tau_{\rm th}}}{{\delta F}\over F},\\
{\delta F\over F} & \sim & {{\delta p}\over p}+\delta s.
\label{eq:nonad-dsdfesti}
\end{eqnarray}
\end{mathletters}

Together, equations
\refnew{eq:nonad-scalesds}-\refnew{eq:nonad-dsdfesti} imply that the
ratio of the nonadiabatic to adiabatic contributions to both $\delta
p/p$ and $\delta F/F$ is of order $(\omega\tau_{\rm th})^{-1}$, as is
commonly cited in the literature.

\subsection{Nonadiabatic Effects in the G-Mode Cavity}
\label{sec:nonad-cavity}

Here, all perturbation quantities vary on a vertical scale $k_z^{-1}$, 
where $k_z z \geq 1$. Moreover, it follows from the local dispersion
relation derived in Paper I that $k_z \sim (z z_\omega)^{-1/2}$.

Scaling the adiabatic terms in equation \refnew{eq:nonad-dppds} yields
\begin{mathletters}
\begin{eqnarray}
{d\over{d\ln p}}\left({{\delta p}\over p}\right) & \sim & (k_z z)
\left({{\delta p}\over p}\right), \\ 
{dX\over{d\ln p}} & \sim & (k_z
z)^2 \left({{\delta p}\over p}\right), \\
\left({p\over{\rho g}}\right)^2
\left[ k_h^2 \left({N^2\over \omega^2} -1\right) + \left({\omega \over
{c^2}}\right)^2 \right] \left({{\delta p}\over p}\right) & \sim & 
{{z}\over{z_\omega}} \left({{\delta p}\over p}\right)
 \sim (k_z z)^2 \left({{\delta p}\over p}\right).
\label{eq:nonad-scalesp2}
\end{eqnarray}
\end{mathletters}
To order of magnitude, the nonadiabatic term is given by
\be
{p\over{\rho g}} {{(gk_h)^2-\omega^4}\over{g \omega^2}} \rho_s
\delta s \sim  {z\over{z_\omega}} \delta s \sim (k_z z)^2 \delta s.
\label{eq:nonad-scalesds2}
\ee
These equations imply that the nonadiabatic correction to $\delta p/p$
is of order $\delta s$.

We scale equations \refnew{eq:nonad-dsdFF} and \refnew{eq:nonad-dffall} to
relate $\delta s$ to $\delta p/p$;
\begin{mathletters}
\begin{eqnarray}
\delta s & \sim & {(k_z z)\over{\omega \tau_{\rm th}}} 
\left({{\delta F}\over F}\right),  \\
{{\delta F}\over F} & \sim & {d\over{d\ln p}}\left({{\delta 
p}\over p}+\delta s\right) \sim (k_z z)\left({{\delta p}\over 
p}+\delta s\right).
\label{eq:nonad-helptwo}
\end{eqnarray}
\end{mathletters}

Combining equations
\refnew{eq:nonad-scalesp2}-\refnew{eq:nonad-helptwo}, we determine
that the ratio of nonadiabatic to adiabatic contributions to $\delta
p/p$ and $\delta F/F$ is of order $(k_z z)^2/\omega\tau_{\rm
th}$. This is not surprising; $\tau_{\rm th}/(k_z z)^2$ is the
timescale of thermal diffusion across distance
$k_z^{-1}$. Nonadiabaticity is measured by the ratio of the mode
period to this timescale.

\section{Solution Of Eigenvalue Problem}
\label{sec:nonad-numer}

We follow a two step procedure in solving the linear
pulsation equations \refnew{eq:nonad-defineX}-\refnew{eq:nonad-dffall}
for the four dependent variables subject to the five boundary
conditions given by equations 
\refnew{eq:nonad-dppzb}-\refnew{eq:nonad-dffzdeep}. The initial step 
consists of guessing a value for the complex eigenfrequency and then
applying a relaxation method (cf. Press \etal \cite{nonad-press92}) to
solve the differential equations subject to four out of the five
boundary conditions.\footnote{The trivial boundary condition given by
equation \refnew{eq:nonad-dppzb} is always included as one of this
foursome.} The step is complete when the dependent variables satisfy
both the pulsation equations and the boundary conditions to $10^{-7}$
of a scaling factor provided by the corresponding adiabatic
eigenfunction at individual points. Working in double precision, this
is routinely achieved. The second step determines the eigenvalue by
requiring the remaining boundary condition to be satisfied.  Both
minimization and root finding techniques work well.  Normally we
reserve equation \refnew{eq:nonad-ddppzdeep} for our fifth boundary
condition, but identical results are obtained when others are used
instead.

\subsection{Eigenvalues for Stellar G-Modes}
\label{subsec:nonad-stellar}

Nonadiabatic eigenvalues of our plane-parallel hydrogen envelopes are
denoted by primes, $\omega^\prime=\omega_r^\prime +i\omega_i^\prime$. 
To relate $\omega_i^\prime$ to $\omega_i$ of a complete DA white dwarf model,
we proceed as follows. We compute adiabatic modes for a corresponding complete 
model and indentify those whose frequencies bracket 
$\omega_r^\prime$.\footnote{The adiabatic 
modes have the same $\ell$ as the mode of the plane-parallel envelope.} We 
determine by interpolation the non-integer number $n$ of nodes in $\delta p/p$ 
to assign to the adiabatic mode of the complete model whose frequency 
$\omega=\omega_r^\prime$. Suppose that $n^\prime$ of these nodes lie above 
$z_{\rm deep}$. Then 
\be
\omega_i\approx {n^\prime\over n}\omega_i^\prime.
\label{eq:nonad-real}
\ee
Validation of equation \refnew{eq:nonad-real} is provided by equation 
\refnew{eq:na-omi} in the appendix. Its physical 
justification is that whereas energy is dissipated near the surface,  it is
stored throughout the mode cavity. 

Our procedure for converting $\omega_i^\prime$ to $\omega_i$ requires
nonadiabatic effects to be small well above $z_{\rm deep}$, as is the
case for all modes of interest to our investigation.  However, it only
applies to modes which have at least one node above $z_{\rm
deep}$. Thus it cannot handle the highest frequency g-modes. Values of
$\omega_i$ for these are computed from the work integral.

Very low order modes ($n \leq 6$) are evanescent above $z_{\rm
deep}$. We therefore cannot enforce the boundary condition equation
\refnew{eq:nonad-ddppzdeep} on them to solve for the eigenvalues. 
We rely on work integral for stability analysis of these modes.

\begin{figure}[t]
\centerline{\psfig{figure=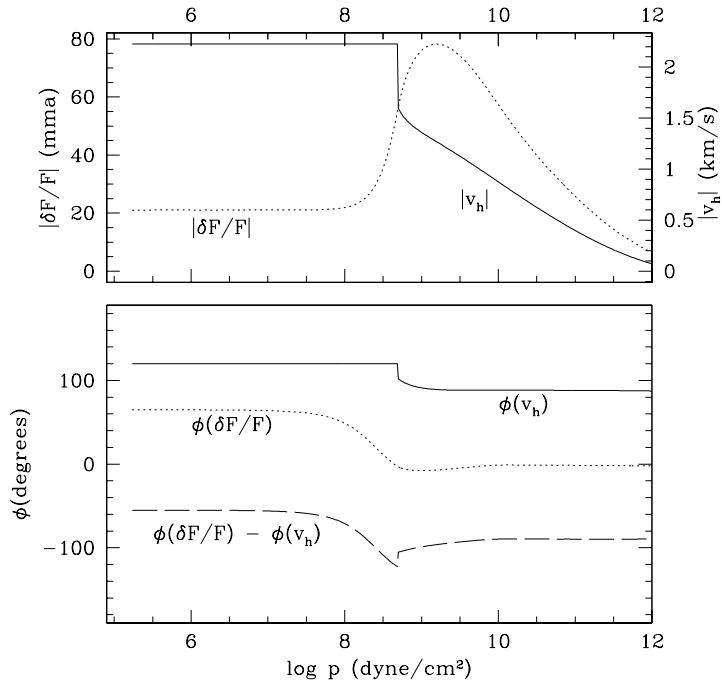,width=0.60\hsize}}
\caption[]{Amplitudes and phases of $\delta F/F$ and $v_h$ as
functions of $\log p$. The displayed mode has $\ell=1$ and $P=800\s$
corresponding to $\omega\tau_{\rm th}=0.8$ at $z=z_b$ in a stellar
model with $T_{\rm eff}=12,\!160\K$, $g=10^8\cm\s^{-2}$, and
$f=0.32$. The amplitude, $|X|$, and phase, $\phi$, of a perturbation
$X$ are defined as $X = |X| \cos(\omega t - \phi)$. In the adiabatic
interior, the flux perturbation leads the velocity perturbation by
$90^\circ$.}
\label{fig:na-nafun4085-800}
\end{figure}

\subsection{The Work Integral}
\label{subsec:nonad-quasi}

Calculating the work integral provides an approximate method for
evaluating the driving or damping rate of a mode. The work integral
computes $\gamma=2\omega_i$. This method is well defined and accurate
when the quasiadiabatic approximation applies. Moreover, it reveals
regions of driving and damping. Since overstable g-modes of ZZ Cetis
have high quality factors, their linear pulsations are nearly
periodic. The work integral may be calculated as follows (Unno \etal
\cite{nonad-unno89}):
\begin{eqnarray}
 \gamma & = &  {{\omega_r}R^2\over {2\pi}} 
\oint dt \int_0^R dz\, \rho\, {{k_B}\over{m_p}} {\delta T}
{{d\delta s}\over{dt}} \nonumber \\ & = & {\omega_r{R^2}\over 2} \int_0^R dz\,
\nabla_{ad} \,\rho \,{{k_B}\over{m_p}}\,T 
 \left[ \left({{\delta p}\over p}\right)_r \delta s_i - 
\left({{\delta p}\over p}\right)_i \delta s_r \right].
\label{eq:nonad-workps}
\end{eqnarray}

\section{Numerical Results}
\label{sec:nonad-result}

\subsection{Nonadiabatic Effects On Eigenfunctions}
\label{subsec:nonad-naquali}

The effects of nonadiabaticity are illustrated by comparing
nonadiabatic and adiabatic eigenfunctions for three $\ell=1$ g-modes
of the same stellar model. The model is characterized by $T_{\rm eff}
= 12,\!000\K$, $g=10^8\cm\s^{-2}$, and $f=0.32$, which together imply
$\tau_{\rm th}\approx 200 \s$ at $z = z_b$. The modes have periods of
$430\s$, $1,\!500\s$, and $2,\!400\s$. Quasiadiabatic calculations predict
overstability for each of these modes.

Radiative diffusion acts to soften sharp temperature gradient
perturbations.  Its importance increases with mode period. These
characteristics are illustrated in Figures
\ref{fig:na-nad430}-\ref{fig:na-nad2400}.\footnote{We normalize
$\delta p/p$ at $z=z_{\rm deep}$ by setting it equal to its adiabatic
counterpart. Thus the imaginary component of $\delta F/F$ is due to
nonadiabaticity.} The $430\s$ mode, with $\omega \tau_{\rm th} = 2.8$
at $z=z_b$, is quite adiabatic in the radiative interior. Radiative
diffusion is more pronounced in the evanescent region of the $1,\!500\s$
mode, which has $\omega\tau_{\rm th}=0.8$ at $z=z_b$. The nonadiabatic
eigenfunction of the $2,\!400\s$ mode, for which $\omega\tau_{\rm
th}=0.5$ at $z=z_b$, deviates significantly from its adiabatic
counterpart in both the evanescent and propagating regions. The phase
shift between $\delta F/F$ and $\delta p/p$ at $z_b$ is so large that
the convection zone contributes to mode damping. This mode is stable.

The recent detection of velocity signals associated with g-modes by van Kerkwijk 
\etal (\cite{nonad-marten98}) is an important advance in the study of ZZ Cetis. 
What is being detected are horizontal velocities near the stellar limb. 
Phase differences and amplitude ratios of flux and velocity variations due to 
individual modes offer unique clues about both the star and the modes.
Nonadiabatic calculations, such as that depicted in Figure 
\ref{fig:na-nafun4085-800}, provide a theoretical basis for interpreting these 
observable quantities. Deep inside the radiative interior, the maximum
flux perturbation preceeds the maximum horizontal velocity by a
quarter of a period, and the flux to velocity amplitude ratio is
determined by the quasiadiabatic approximation. These relations are
modified in the upper envelope by the convective transport of heat and
momentum and by radiative diffusion. Convection delays and diminishes
the flux variation (see Paper I). It also flattens the horizontal
velocity profile above $z_b$ and forces it to jump at $z_b$ (see Paper
II). Radiative diffusion smears flux perturbations below the
convection zone. In the example illustrated in Figure
\ref{fig:na-nafun4085-800}, a photospheric flux variation of
$20$\,mma\footnote{Milli-magnitude-amplitude is the commonly adopted
unit for measuring flux perturbations. An $1$\, mma variation means
$\delta F/F=10^{-3}$.}  corresponds to a horizontal velocity of
$v_h=2.3 \km\s^{-1}$.  Nonadiabaticity, mostly in the convection zone,
reduces the phase lag between velocity maximum and light maximum from
the quasiadiabatic value of $90^\circ$ to $55^\circ$ at the
photosphere. Employing the conventions in van Kerkwijk \etal
(\cite{nonad-marten98}), this mode would exhibit $R_v = |v_h|/(\omega
|\delta F/F|) =16$, and $\Delta \Phi_v = \Phi(\delta F/F) - \Phi(v_h)=
-55^\circ$. This is consistent with the $818\s$ mode which they
observed to have $R_v = 11\pm 4$ and $\Delta \Phi_v = -44^\circ\pm
19^\circ$.\footnote{Effects of limb-darkening, disc averaging, and
bolometric correction introduce a factor close to unity.}

\subsection{Nonadiabatic Effects On Driving and Damping Rates}
\label{subsec:nonad-naquant}

\begin{figure}[t]
\centerline{\psfig{figure=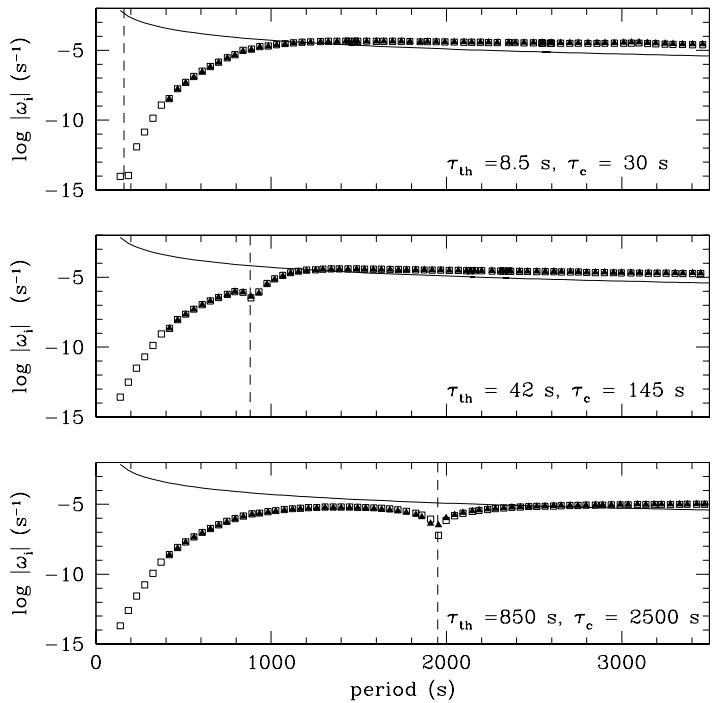,width=0.6\hsize}}
\caption[]{Driving and damping rates for ${\ell} = 1$ modes plotted 
against mode periods. Top to bottom, the panels pertain to stellar
models computed with $g=10^8\cm\s^{-1}$ and $f=0.32$ for $T_{\rm eff}
= 12,\!600 \K$, $12,\!300 \K$, and $11,\!750 \K$. Values of
$|\omega_i|$ obtained from the nonadiabatic code are denoted by solid
triangles. Those calculated from work integrals using nonadiabatic
eigenfunctions are shown by open squares.  For very short periods, $P<
400 \s$, our shallow envelopes force us to rely entirely on the work
integral. The dashed vertical line marks the boundary between shorter
period overstable modes and longer period damped ones. The solid line
displays the analytic estimate, $\omega_i=-\omega_r\ln{\cal
R}^{-1}/(2\pi n)$, for the damping rate of modes which suffer strong
dissipation.  We set ${\cal R} = 1/e$ here, although in reality it
decreases with increasing mode period and effective temperature.}
\label{fig:na-nonadratecombi}
\end{figure}

\begin{figure}[t]
\centerline{\psfig{figure=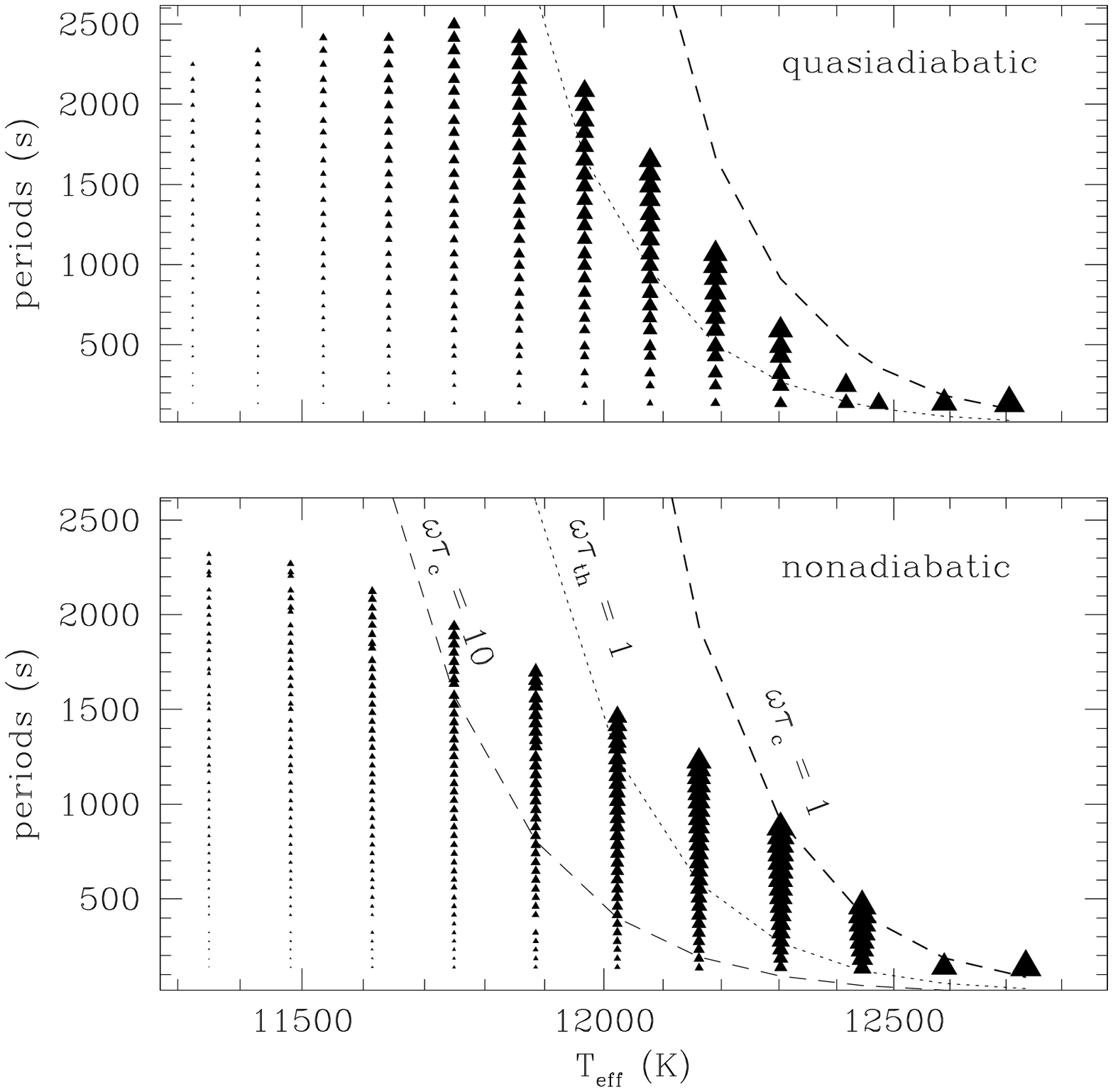,width=0.60\hsize}}
\caption[]{Panoply of overstable $\ell=1$ modes according to quasiadiabatic
(upper panel) and fully nonadiabatic (lower panel) calculations in
white dwarf models produced with $g=10^8\cm\s^{-1}$ and $f=0.32$ for a
range of $T_{\rm eff}$ which covers the ZZ Ceti instability
strip. Each overstable mode is marked by a solid triangle whose size
is proportional to the diminution factor of the flux perturbation in
the convection zone, $[1+(\omega \tau_c)^2]^{-1/2}$ (see Paper I).
Dashed lines denote the loci of constant values of $\omega\tau_c$.
The dotted line corresponds to $\omega \tau_{\rm th} = 1$ with
$\tau_{\rm th}$ evaluated at $z=z_b$. }
\label{fig:na-vis-jumpglobal} 
\end{figure}

Where applicable, direct calculations of $\omega_i$ yield
results consistent with those based on the work integral. This
provides a measure of confidence in both.  Modest discrepancies are
found for some marginally overstable modes. The eigenvalue
calculations are more reliable, as work integrals suffer from
inaccuracies due to 
cancellation between comparable magnitudes of driving and damping.
Values of $\omega_i$ for individual g-modes evolve with decreasing
$T_{\rm eff}$. Effects of nonadiabaticity on driving and damping rates
of g-modes are presented in Figures
\ref{fig:na-nonadratecombi}-\ref{fig:na-vis-jumpglobal}. General
trends are described below.

Upper lids of cavities of short period modes (e.g., the $430\s$ mode) lie far 
beneath the bottom of the convection zone in even in the coolest variables. 
Their driving/damping rates are largely immune to both nonadiabatic effects and 
the depth of the convection zone, and hardly vary across the instability strip.  
The quasiadiabatic estimate, $\omega_i
\sim 1/(n\tau_\omega)$, pertains to these modes (see Paper I).

Longer period modes (e.g., the $1,\!000 \s$ mode) become overstable at
lower $T_{\rm eff}$.  Values of their $|\omega_i|$ exhibit a
steady decline with decreasing $T_{\rm eff}$. This is a consequence of
the increase in mode inertia with decreasing $T_{\rm eff}$. As the convection 
zone deepens, it depresses the upper lid of a mode's cavity. This decreases 
the relative size of the perturbation amplitude 
near the surface, where nonadiabaticity is greatest, with respect to that 
in the interior, where most of the mode energy is stored. 

Modes with even longer periods (e.g., the $2,\!000 \s$ mode) are weakly
overstable in cooler and narrower temperature ranges. Nonadiabatic
effects tend to stabilize these modes. In the limit of strong
dissipation, $\omega_i/\omega_r\approx -\ln{{\cal R}^{-1}}/(2\pi n)$,
where ${\cal R}$ is the reflection coefficient at the cavity lid. We
derive this relation using a toy model in the appendix.  Magnitudes of
nonadiabatic damping rates are an important input to calculations of
parametric instability, an amplitude limiting mechanism for overstable
modes that is explored in the next paper of this series.

\section{Maximum Periods of Overstable Modes}
\label{sec:nonad-maximum}

\subsection{Limitations of the Overstability Criterion $\omega\tau_c>1$}

The derivation of the overstability criterion $\omega \tau_c> 1$ depends on
the quasiadiabatic approximation and the condition that $z_\omega \gg z_b$ 
(Brickhill \cite{nonad-brick91}, Paper I). Since $\tau_c\approx 4\tau_{\rm th}$ 
at $z_b$, the quasiadiabatic approximation is suspect for modes with $\omega 
\tau_c\approx 1$. The upper and lower panels of Figure 
\ref{fig:na-vis-jumpglobal} display the boundary between overstable and damped 
modes as determined by quasiadiabatic and nonadiabatic
calculations. Both show that the boundary value of $\omega \tau_c$ is
close to unity for hot ZZ Cetis but rises as stars cool. Inaccuracy of
the quasiadiabatic approximation does not dramatically modify the
overstability criterion, but the violation of the condition
$z_\omega\gg z_b$ does. The latter occurs because $z_b$ increases with
decreasing $T_{\rm eff}$ and $z_\omega$ decreases with decreasing
$\omega$.  Modes with $z_\omega\lesssim z_b$ propagate just below
$z_b$.  Their short WKB wavelengths enhance radiative damping above
the estimate given by equation (54) in Paper I on which the derivation
of the overstability criterion $\omega \tau_c> 1$ rests.

\subsection{Comparison with Observations}

\begin{figure}[t]
\centerline{\psfig{figure=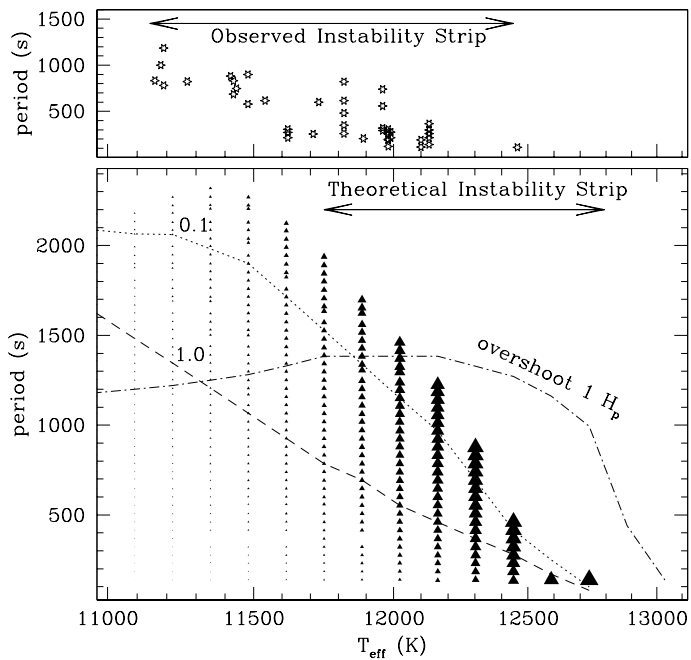,width=0.60\hsize}}
\caption[]{Observational and theoretical instability strips for DA white
dwarfs. The upper panel displays each of the known ZZ Cetis according
to the dominant period in its pulsation spectrum (Bradley 
\cite{nonad-bradley95}), and its inferred effective temperature (Bergeron
\etal \cite{nonad-bergeron95}).  The lower panel repeats the material 
shown in the lower panel of Figure \ref{fig:na-vis-jumpglobal}. The
dotted and dashed lines illustrate the maximum period for overstable
modes under the stabilizing effect of turbulent damping in the region
of convective overshoot for values of the parameter $\lambda/z_b$
equal to 0.1 and 1.0, respectively.  The dotted-dashed line shows the
envelope of overstable modes for models which incorporate one pressure
scale height of convective overshoot.  When comparing observational
and theoretical instability strips, one should bear in mind that both
depend upon the assumed mixing-length.}
\label{fig:na-vis-L1mode-new} 
\end{figure}

Both our quasiadiabatic and nonadiabatic calculations 
consistently predict maximum periods for overstable modes that are
about a factor two longer than those observed
(cf. Fig. \ref{fig:na-vis-L1mode-new}). We have not been able to
resolve this discrepancy. Several possible explanations which we have
considered are described in what follows.

\subsubsection{Effects of Turbulent Damping}
\label{subsubsec:nonad-turbulent}

Both Brickhill (\cite{nonad-brick90}) and Paper II stress that
linear damping due to turbulent viscosity, although negligible inside the 
convection zone for all modes, may be significant in the region of convective 
overshoot for long period modes. We incorporate this effect into our analysis, 
while recognizing that it cannot be quantified precisely.

The rate of turbulent damping in the overshoot region is expressed as
(eq. [36] in Paper II),
\be
\gamma_{\rm vis-os} \approx {-\pi R^2\rho_b\omega^3\lambda
 |\Delta\xi_h|^2\over 4},
\label{eq:na-vis-os}
\ee
where $|\Delta \xi_h|$ is the normalized jump in the horizontal displacement 
across the convective overshoot region within which the turbulent 
viscosity decays on length scale $\lambda$. To cover our ignorance, we 
present results for $\lambda/z_b$ of $0.1$ and $1.0$. Figure 
\ref{fig:na-vis-L1mode-new}
demonstrates that with the larger value, the longest period overstable
mode at a given $T_{\rm eff}$ is compatible with the observed upper
limit of $P=1,\!200\s$.  However, it seems unlikely that
$\lambda/z_b\approx 1$. The overshoot region probably extends less
than a pressure scale height below $z_b$, while $H_p\approx 0.4z_b$,
and $\lambda$ must be several times smaller than $H_p$.

\subsubsection{Effects of Convective Overshoot}
\label{subsubsec:nonad-overshoot}

We modify our envelope models to simulate convective overshoot by
adding an isentropic layer below the base of the convection
zone. Within this layer the radiative flux exceeds the total flux and
turbulence transports energy downward. 
The top of the radiative interior immediately below the overshoot layer
differs from that below a convective layer of similar depth. The temperature
gradient is shallower and the \Bruntfreq frequency has a finite positive value.
These differences enhance the effects of radiative 
diffusion and modify the relations between horizontal velocity and
light perturbations at the photosphere.

\begin{figure}[t]
\centerline{\psfig{figure=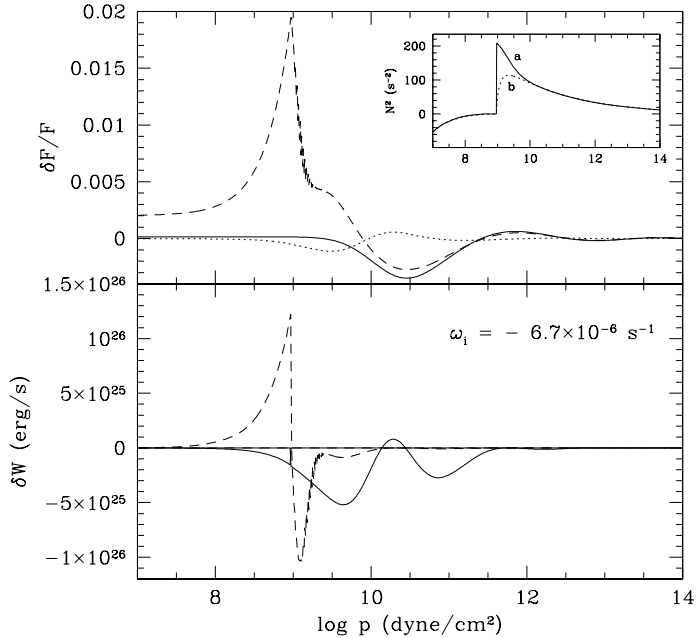,width=0.6\hsize}}
\caption[]{Effects of convective overshoot on g-mode structure and stability. 
The stellar model lies at $T_{\rm eff} = 12,\!250 \K$ with one pressure
scale height of overshoot below the convection zone.  We use the same
mode as in Figure \ref{fig:na-nad1500-new}, and the stellar model
there has a convection zone as deep as the bottom of the overshoot
region in this model.  The small inset in the top panel shows the
\Bruntfreq frequencies for the former in dashed line (labelled with
'b') and the latter in solid line (labelled with 'a').  The top and
bottom panels display similar materials as in Figure
\ref{fig:na-nad1500-new}, except that the nonadiabatic work integral is
multiplied by a factor of $20$ to aid display. Both quasiadiabatic and
nonadiabatic calculations predict stability for this mode. }
\label{fig:na-nad1500-os}
\end{figure}

Figure \ref{fig:na-nad1500-os} provides such an example.  Our
nonadiabatic calculations indicate that one pressure scale height of
overshoot suffices to stabilize all modes with $P > 1,\!400 \s$ (see
Fig. \ref{fig:na-vis-L1mode-new}).  Moreover, by effectively deepening
the convection zone, it shifts the instability strip to higher $T_{\rm
eff}$.

\subsubsection{Sensitivity to Surface Gravity}
\label{subsubsec:nonad-gravity}

Our nonadiabatic calculations indicate that, at fixed $T_{\rm eff}$,
the longest period for an overstable mode scales approximately as
$g^{-1/3}$.  Thus maximum periods of overstable modes around $1,\!200
\s$ would require $g \approx 5.0 \times 10^{8} \cm \s^{-2}$, which is
well outside observational constraints.\footnote{Applying a
theoretical mass-radius relation (Hamada \& Salpeter
\cite{nonad-salpeter61}) for white dwarfs, this corresponds to a
stellar mass about $1.2 M_\odot$.}

\subsubsection{Other Observational and Theoretical Considerations}
\label{subsubsec:nonad-bias}

Hansen \etal (\cite{nonad-hansen85}) show that damping due to upward
propagation of a gravity wave above the photosphere is unimportant for
modes having frequencies exhibited ZZ Cetis. We concur with this
conclusion.

Detection of lower frequency modes requires longer observational data
streams. Noise due to variations of atmospheric transparency increases
at lower frequencies (see Fig. 4 of Winget
\cite{nonad-winget91}). These may result in an observational bias
against the detection of low frequency modes.  

Evidence that photometric amplitudes of modes decline for $P \gtrsim
1,\!000 \s$, suggests that the observational cutoff at $P \approx 1,\!200
\s$) is genuine (see Fig. 5 in Clemens \cite{nonad-clemens95}). Taken
literally, it also hints that the cutoff is the result of a nonlinear
mechanism which saturates pulsation amplitudes.

\section{Summary}
\label{sec:nonad-sum}

There is little doubt that convective driving, as originally proposed
by Brickhill (\cite{nonad-brick91}), is responsible for the linear
overstability of g-modes in DA white dwarfs.  It is physically
self-consistent, and convincingly rationalizes observational facts. It
accounts for the general location of the instability strip, although a
precise specification depends upon the modeling of convection (e.g.,
the mixing-length parameter). Convective driving also explains why
longer period modes become overstable as a star cools (see Fig.
\ref{fig:na-vis-L1mode-new}). 

Although we agree with Brickhill that $\omega \tau_c > 1$ is a
necessary condition for mode overstability, we find that it is not a
sufficient condition for modes whose periods exceed $1,\!000\s$. This
stems from enhanced radiative damping of modes whose upper cavity lids
approach $z_b$, as is apparent from both quasiadiabatic and
nonadiabatic calculations (see Fig. \ref{fig:na-vis-jumpglobal}). Our
nonadiabatic calculations yield a maximum period of about $2,\!300\s$
for overstable modes. This clashes with the maximum period of
$1,\!200\s$ for observationally detected modes.

We also agree with Brickhill's deduction (\cite{nonad-brick90}) that
turbulent convection forces the horizontal velocity to be nearly
independent of depth within the convective envelope. Consequently,
mode damping due to turbulent dissipation within the convection zone
is reduced to a negligible level. However, suppression of the
horizontal shear in the convective envelope results in a shear layer
at the top of the radiative interior. When convective overshoot is
accounted for, this provides linear turbulent damping. Figure
\ref{fig:na-vis-L1mode-new} suggests that turbulent dissipation might
depress the maximum period of observable modes to some degree.
Convective overshoot also alters the thermal structure of the upper
radiative layer. A rough treatment of this effect with one scale
height of overshoot predicts a maximum period of $1,\!400 \s$ for
overstable modes. Nonlinear interactions which limit the amplitudes of
overstable modes may also play a part in determining the maximum
period.

In conclusion, the blue edge of the theoretical instability strip
seems to be set by the condition that $\omega\tau_c\approx 1$ for the
lowest order $\ell=1$ mode. But the location of the red edge is more
nebulous, and may result from a combination of decreased photometric
visibility, convective overshoot, and nonlinear effects.  Additional
detections of velocity variations associated with g-modes could
provide important clues. Convective driving makes the testable
prediction that velocity variations become relatively more observable
than photometric variations toward the red edge of the instability
strip.

\begin{appendix}
\section{{\it A Toy Model For Nonadiabatic Modes}}
\label{subsubsec:nonad-onedimen}

We describe a simple toy model for nonadiabatic modes. It is
particularly useful for interpreting damping rates in the limit of
strong dissipation.

Consider waves which satisfy the one-dimensional, homogeneous,
acoustic wave equation
\be
{{\partial^2 \xi}\over {\partial t^2}} =  u^2 {{\partial^2
\xi}\over{\partial z^2}},
\label{eq:na-homowave}
\ee
in the interval $0\leq z\leq L$. Here $\xi$ is the Lagrangian displacement and
$u$ is the constant propagation speed. The dispersion relation connecting
frequency, $\omega$, and wave vector, $k$, reads
$\omega^2=k^2u^2$. We take the lower boundary to be a rigid, perfectly 
reflecting wall, so
\be
\xi=0 \quad {\rm at} \quad z=L. 
\label{eq:na-bdcL}
\ee
Dissipation is introduced by means of a partially reflective upper
boundary, where ${\cal R}$ denotes the reflection coefficient of the incident 
wave. This is expressed through the boundary condition
\be
\left({\partial\over\partial t}-u{\partial\over\partial z}\right)\xi
= -{\cal R}\left({\partial\over\partial t}+u{\partial\over\partial 
z}\right)\xi \quad {\rm at} \quad z=0.
\label{eq:na-bdcU}
\ee
Note that in the limit ${\cal R}\to 1$, the upper boundary becomes a perfectly 
reflecting wall.  

Eigen-solutions of equation \refnew{eq:na-homowave} are composed of 
oppositely directed waves;
\be
\xi = A e^{-i\omega t - i k z} + B e^{-i \omega t + i k z}.
\label{eq:na-homowavesol}
\ee
Application of the boundary conditions given by equations
\refnew{eq:na-bdcL} and \refnew{eq:na-bdcU} yields
\be
B=-{\cal R}A,
\label{eq:na-reflect}
\ee
and
\be
kL=n\pi-{i\over 2}\ln{\cal R}^{-1},
\label{eq:na-k}
\ee
where $n$ is the number of half wavelengths between the walls. Then the 
dispersion relation implies  
\be
\omega_i=
{{k_i}\over{k_r}} \omega_r = 
-{\omega_r\over 2\pi n}\ln{\cal R}^{-1},
\label{eq:na-omi}
\ee
where $n$ is an integer. 

Equation \refnew{eq:na-omi} is the key result of our toy model. It demonstrates 
that $\omega_i$ grows logarithmically with ${\cal R}^{-1}$, and provides 
an order-of-magnitude estimate for the damping rate of strongly nonadiabatic 
stellar modes. Radiative diffusion substantially suppresses the effective 
reflection coefficients of modes having $\omega \tau_{\omega}
\leq 1$. In hot DA stars, this applies to modes with periods in excess of
$1,\!000\s$ (cf. Fig. \ref{fig:na-nonadratecombi}). Finally, equation 
\refnew{eq:na-omi} justifies the procedure we apply in \S 
\ref{subsec:nonad-stellar} to translate $\omega_i^\prime$, the damping rate for 
a mode of an envelope model, to $\omega_i$, the damping rate for a mode of a 
complete stellar model.

\end{appendix}

\end{document}